\documentclass[twocolumn]{aastex63}
\submitjournal{ApJL}
\shorttitle{Gamma-suppression in neutrino-blazars}
\shortauthors{Kun et al.}
\usepackage{fancyvrb} 
\usepackage{breakurl}
\usepackage{gensymb}
\VerbatimFootnotes    

\begin{document}

\title{Cosmic neutrinos from temporarily gamma-suppressed blazars}

\correspondingauthor{Emma Kun}
\email{kun.emma@csfk.org}

\author[00000-0003-2769-3591]{Emma Kun}
\affiliation{Konkoly Observatory, Research Centre for Astronomy and Earth Sciences, Eötvös Loránd Research Network (ELKH)\\
H-1121 Budapest, Konkoly Thege Miklós út 15-17., Hungary}
\author[0000-0001-5607-3637]{Imre Bartos}
\affiliation{Department of Physics, University of Florida, PO Box 118440, Gainesville, FL 32611-8440, USA}
\author[0000-0002-1748-7367]{Julia Becker Tjus}
\affiliation{Theoretical Physics IV: Plasma-Astroparticle Physics, Faculty for Physics \& Astronomy, Ruhr-Universit\"at Bochum, 44780 Bochum, Germany}
\affiliation{Ruhr Astroparticle And Plasma Physics Center (RAPP Center), Ruhr-Universit\"at Bochum, 44780 Bochum, Germany}
\author[00000-0000-0000-0000]{Peter L.\ Biermann}
\affiliation{MPI for Radioastronomy, 53121 Bonn, Germany}
\affiliation{Department of Physics \& Astronomy, University of Alabama, Tuscaloosa, AL 35487, USA}
\author[0000-0001-6224-2417]{Francis Halzen}
\affiliation{Dept. of Physics, University of Wisconsin, Madison, WI 53706, USA}
\author[0000-0002-0686-7479]{Gy\"orgy Mez\H o}
\affiliation{Konkoly Observatory, Research Centre for Astronomy and Earth Sciences, Eötvös Loránd Research Network (ELKH)\\
H-1121 Budapest, Konkoly Thege Miklós út 15-17., Hungary}

\begin{abstract}
Despite the uncovered association of a high-energy neutrino with the apparent flaring state of blazar TXS\,0506+056 in 2017, the mechanisms leading to astrophysical particle acceleration and neutrino production are still uncertain. Recent studies found that when transparent to $\gamma$-rays, $\gamma$-flaring blazars do not have the opacity for protons to produce neutrinos. Here we present observational evidence for an alternative explanation, in which $\gamma$-ray emission is suppressed during efficient neutrino production. A large proton and target photon density help produce neutrinos while temporarily suppress the observable $\gamma$-emission due to a large $\gamma \gamma$ opacity. We show that the Fermi-LAT $\gamma$-flux of blazar PKS\,1502+106 was at a local minimum when IceCube recorded the coincident high-energy neutrino IC-190730A.  Using data from the OVRO 40-meter Telescope, we find that radio emission from PKS\,1502+106 at the time period of the coincident neutrino IC-190730A was in a high state, in contrast to earlier time periods when radio and $\gamma$ fluxes are correlated for both low and high states. This points to an active outflow that is $\gamma$-suppressed at the time of neutrino production. We find similar local $\gamma$-suppression in other blazars, including in MAGIC's TeV flux of TXS\,0506+056 and Fermi-LAT's flux of blazar PKS\,B1424–418 at the time of coincident IceCube neutrino detections. Using temporary $\gamma$-suppression, neutrino-blazar coincidence searches could be substantially more sensitive than previously assumed, enabling the identification of the origin of IceCube's diffuse neutrino flux possibly with already existing data.  
\end{abstract}

\keywords{galaxies: active, galaxies: individual (PKS 1502+106), gamma rays: galaxies, neutrinos,  radio continuum: galaxies}

\section{Introduction}
\label{section:intro}

The origin of the isotropic cosmic flux of high-energy neutrinos, discovered by IceCube in 2013 \citep{Aartsen2013Science,Aartsen2013PRL,Aartsen2020}, is still unknown. The most promising strategy to identify the underlying sources has been multi-messenger observations \citep{ICTXS2018a,Murase2019}. Studies aiming to find sources of cosmic neutrinos have recently focused on the blazar subclass of active galactic nuclei \citep[AGN, e.g.][]{Kadler2016,Krauss2018,Franckowiak2020,Giommi2020,Hoerbe2020}, extreme particle accelerators directing their relativistic $\gamma$-ray jet toward Earth. 

The spectral energy distribution of blazars is characterized by a low and a high-energy bump \citep[e.g.][]{Abdo2010}. The low-energy component is due to synchrotron radiation of charged particles gyrating about the magnetic field of the AGN, which is the primary process generating the radio continuum of jetted AGN \citep[e.g.][]{Boettcher2012}. The origin of the high-energy emission is less clear, as the observed spectrum can be explained by leptonic, hadronic or hybrid lepto-hadronic models \citep[e.g.][]{Murase2018,Rodrigues2019,Gao2019}. In leptonic models the synchrotron self-Compton process or external inverse Compton fields, while in hadronic models synchrotron emission of high-energy photons and the decay of neutral pions can lead to the enhanced $\gamma$-ray emission of blazars \citep[e.g.][]{Biermann2011,Aharonian2013}. Above 100 GeV, the emitted radiation is attenuated by Extragalactic Background Light \citep[e.g.][]{Gilmore2012}.

A multimessenger campaign initiated by the observation of a muon neutrino of $290$~TeV energy (IC-170922A) identified the $\gamma$-ray flaring blazar TXS 0506+056 \citep{ICTXS2018a} as source of the neutrino. It was argued that the rapid flux density variation of TXS\,0506+056 at the arrival time of neutrino came from the VLBI core at 15~GHz \citep{Kun2019} and at 43~GHz \citep{Ros2020}. Knowing where to look, IceCube retrieved a second neutrino flare from the archival data producing $13 \pm 5$ neutrinos over $158$~days between 2014 and 2015 \citep{ICTXS2018b}. The flare dominates the integrated flux of $10$ years of IceCube data, leaving the flare containing IC-170922A as a distant second. In contrast to IC-170922A, the $\gamma$-activity of the source was low at the time of the 2014/2015 neutrino flare \citep{Garrappa2019}. \citet{Halzen2019} suggested this could be the result of large optical depths for photomeson production required by efficient high-energy neutrino production, which in turn prevent the escape of high-energy pionic $\gamma$-rays. A population of $\gamma$-ray dark cosmic ray accelerators as a neutrino origin was already suspected \citep[e.g.][]{Murase2016}.

IceCube recorded its highest-energy cosmic neutrino alert ever on 2019 July 30, IC-190730A \citep{IceCube190730A}, triggered by a muon neutrino with an estimated energy of $300\,\mathrm{TeV}$. The Fermi-LAT strong point source 4FGL~J1504.4+1029, associated with the active galaxy PKS~1502+106 \citep[$z=1.838$,][]{Paris2017}, is located within $0.31$ degrees from the best-fit neutrino location. 4FGL~J1504.4+1029 is one of the most brightest $\gamma$-ray sources. Given the large redshift of $1.844$, the this source possesses an extremely high intrinsic luminosity. \citet{Kiehlmann2019} reported that the flux density of PKS~1502+106 at 15~GHz measured with the OVRO 40~m Telescope \citep{Richards2011} shows a long-term outburst peaking at the detection time of the neutrino. The VLBI flux density of PKS\,1502+106 is dominated by the core at 15, 43 and 86\,GHz observing frequencies \citep{Karamanavis2016}. \citet{Xavier2020} found three different activity states, one quiescent, and two flaring states distinguished by the hardness of the X-ray spectra. Recently \citet{Britzen2021} found evidence for an evolving radio ring structure in PKS 1502+106. They also found that the radio emission is correlated with the $\gamma$-ray emission, with radio lagging the $\gamma$-rays. They suggested that the neutrino is most likely produced by $pp$ interaction in the blazar zone.

In this {\it Letter} we examine the temporal structure  of $\gamma$-ray emission for blazars TXS\,0506+056, PKS\,1502+106 and PKS B1424–418, which have associated high-energy neutrino counterparts. For 4FGL~J1504.4+1029 (PKS~1502+106) we analyze the Fermi-LAT light curve and its correlation with the single dish radio flux density curve obtained by the OVRO 40~m Telescope \citep{Richards2011}, and emission properties at the time of the neutrino detected by IceCube. For each of the three blazars we examine, we find that $\gamma$-emission is locally suppressed at the time of the detection of their neutrino counterparts, and rapidly rises in the following weeks. Based on neutrino, $\gamma$-ray and radio observations we present a scenario in which the neutrino emission is expected in temporary $\gamma$-suppressed states of jetted AGN.

\begin{figure*}
\centering
\includegraphics[scale=0.7,angle=270]{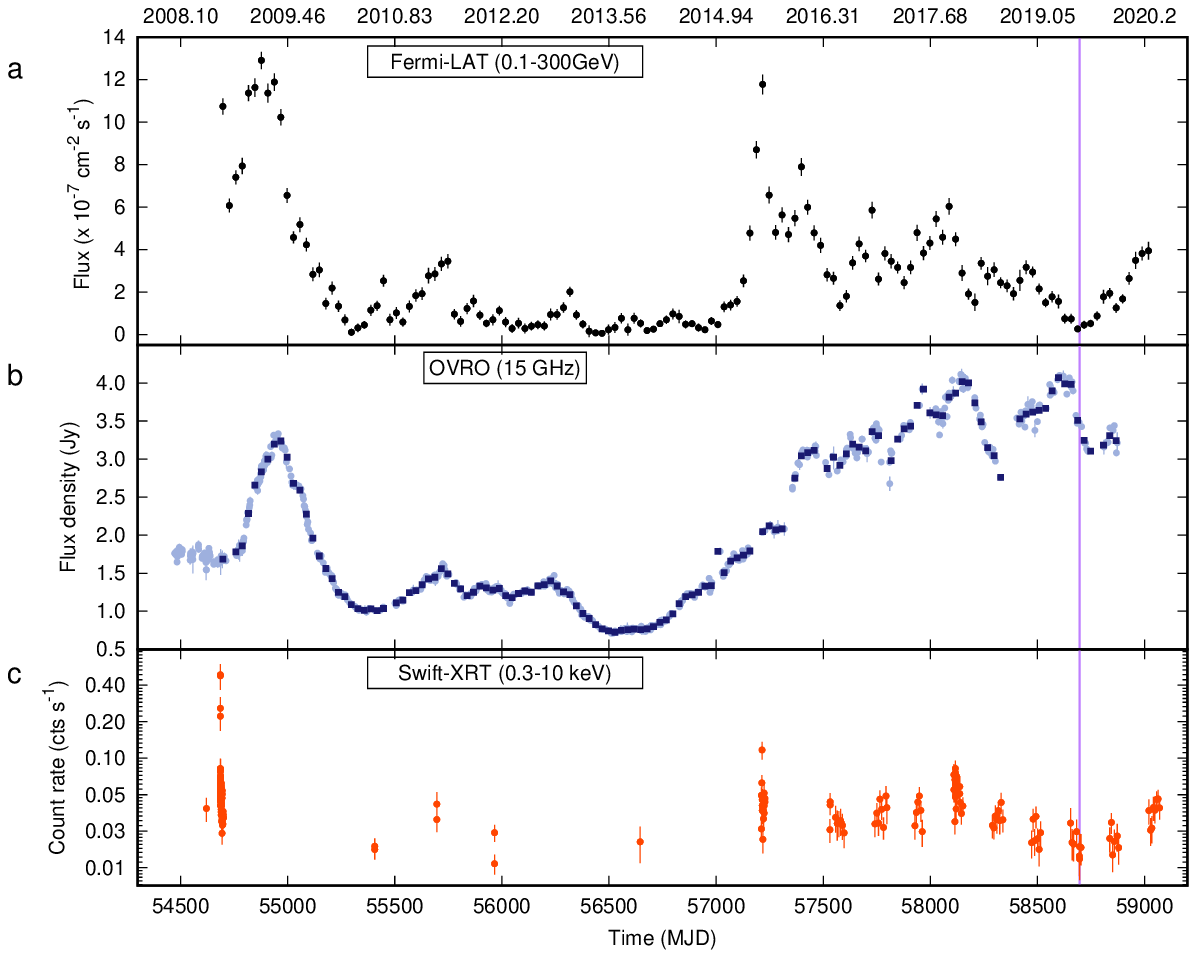}
\caption{{\bf Temporal variation of the $\gamma$-ray, radio, and X-ray brightness of PKS\,1502+106.} {\bf a}: Fermi-LAT likelihood light curve integrated between 100\,MeV and 300\,GeV (recent work, marked by black dots with error bars). {\bf b}: Archive OVRO flux density curve of PKS\,1502+106 plotted with light blue dots, that is superimposed by the radio flux density curve binned to the Fermi-LAT light curve (marked with dark blue squares). {\bf c}: Archive Swift-XRT light curve between 0.3\,keV and 10.0\,keV (marked by red dots with error bars). The detection time of the neutrino IC-190730A is labeled by a vertical purple line. \label{fig:lightcurves}}
\end{figure*}

\section{Fermi-LAT analysis of PKS\,1502+106} 

We analyzed 12 years of Fermi-LAT\footnote{\burl{https://fermi.gsfc.nasa.gov/science/instruments/lat.html}} Pass8 data\footnote{\burl{https://fermi.gsfc.nasa.gov/ssc/data/analysis/documentation/Pass8_usage.html}} around the sky position of 4FGL J1504.4+1029 (PKS 1502+106, see also \citet{Franckowiak2020}), obtained between 2008 August 4 and 2020 June 26 in the energy range from 100\,MeV to 300\,GeV. The search window was centered at $\mathrm{RA_{J2000}}={226.10}\degree$ and $\mathrm{DEC_{J2000}}={10.49}\degree$ encompassing an area of the sky within $\degree{15}$ ROI radius around the central coordinates. We selected event type "front+back" (evtype=3) which is the recommended type for a point source ana{\-}lysis. We performed an unbinned likelihood analysis chain of the data utilizing the Fermi-LAT ScienceTools package v11r5p3 with fermipy package 0.17.3. The instrument response  function \verb+P8R3_SOURCE_V2+ was employed altogether with templates of the Galactic interstellar emission model \verb+gll_iem_v07.fits+ and of the isotropic diffuse emission \verb+iso_P8R3_SOURCE_V2_v1.txt+\footnote{\burl{https://fermi.gsfc.nasa.gov/ssc/data/access/lat/BackgroundModels.html}}. We applied the nominal data quality cut \verb+(DATA_QUAL > 0) && (LAT_CONFIG==1)+, and a zenith angle cut $\theta<{90}\degree$ to eliminate Earth limb events.

We applied a 30-days wide bin width to generate the $\gamma$ light curve of PKS~\,1502+106 which is an unusually bright $\gamma$-ray source dominating its environment throughout the 12 years. Therefore the input model for the light-curve generation included only PKS\,1502+106 (both the prefactor and index were free parameters of the likelihood models)
and diffuse components.

We show the $\gamma$-ray light curve of PKS\,1502+106 in the top panel of Fig.~\ref{fig:lightcurves}. The blazar exhibited larger and smaller flares throughout the Fermi observations. Notably, at the time of the detection of a coincident neutrino neutrino, its $\gamma$--flux was at a local minimum, showing rapid rise during the weeks following the minimum.

\section{Correlation of $\gamma$ and radio emissions of PKS 1502+106}

We show the flux density curve of PKS\,1502+106 obtained by the OVRO 40~m Telescope \citep{Richards2011} at $15$\,GHz in the second panel of Fig.\ \ref{fig:lightcurves}, such that the radio flux densities are binned to the Fermi light curve. It is apparent that the $\gamma$ flux and radio flux density changed simultaneously until about the end of 2014, both for high and for low flux states, and their derivatives changed similarly (mode1). After 2014, however, the radio flux density increased and stayed high while the $\gamma$ flux generally decreased after a sharp rise in 2015 (mode2), such that both $\gamma$ and radio changed more violently and less synchronized. 

We calculated the Pearson correlation index $R$ between the $\gamma$-ray light curve and the binned radio flux density curve in mode1. Its resulted value $R=0.85$ reveals a strong, positive linear correlation. We cross-correlated the $\gamma$-ray light curve with the unbinned radio flux density curve in mode~1 by applying a a python implementation \citep{Robertson2015} of the Discrete Correlation Function method \citep[DCF,][]{Edelson1988}. The DCFs with $15$, $30$, $60$, $90$ and $120$ days wide bins followed similarly shaped distributions centered at about $\tau \approx 60$ days. We estimated the centroid of the DCF, $\tau_c=(\Sigma_i \tau_i \mathrm{DCF}_i)/(\Sigma_i \mathrm{DCF}_i)$ \citep[e.g.][]{Raiteri2011} to be $\tau_c\approx59$ days for the 60-day binning. The uncertainty of $\tau_c$ was estimated by performing Monte Carlo simulations following the “flux redistribution/random subset selection” technique \citep[FR/RSS,][]{Peterson1998}. 
Time lag with uncertainties emerged at $\tau_c=59\pm 32$ days, pointing toward the possibility that the radio flux density curve is delayed with respect to the $\gamma$-ray light curve. Taking into account the whole flux curves, \citep{Britzen2021} found a lag of $151_{-95}^{+40}$ days, similarly the radio lagging behind the $\gamma$-flux.

In mode$2$ the $\gamma$-ray flux was generally decreasing after a rapid increase in the first half of 2015, while the radio flux density was at high state, suggesting a more complex behavior. 

\section{$\gamma$-suppression in neutrino-blazars}
\label{section:scenario}
\subsection{A multimessenger scenario}

In hadronic blazar models the photomeson production is typically assumed to generate high-energy neutrinos in $p\gamma$ interactions because the low particle density in the jet renders the $pp$ process subdominant \citep[e.g.][]{Boettcher2012}. The interaction of cosmic rays in the jet with stellar envelopes or gas clouds \citep[e.g.][]{Barkov2010} might, however, generate high-energy neutrinos through the $pp$ process.

In the followings we assume that cosmic rays are proton dominated, since photo-disintegration of heavier nuclei is unable to provide an effective mechanism for very high energy $\gamma$-ray emission from astrophysical sources \citep{Aharonian2010}.

The flavor-averaged diffuse high-energy cosmic neutrino flux is related to cosmic ray flux by \citep{Halzen2019}:
\begin{equation}
    \frac{1}{3} \sum_\alpha E_\nu ^2 \frac{dN_\nu}{dE_\nu} \simeq \frac{c}{4\pi} \left( \frac{1}{2} (1-e^{-\tau_{p\gamma}}) \xi_z t_H \frac{dE}{dt}\right),
\end{equation}
where the factor $\xi_z$ accounts for the redshift evolution of sources \citep{Ahlers2018}. The IceCube all-flavor diffuse neutrino flux $\sim 3\times 10^{-11}\,\mathrm{TeV}\,\mathrm{cm}^{-2}\,\mathrm{s}^{-1}\, \mathrm{sr}^{-1}$ and the injection rate of cosmic rays above $10^{16}$ eV $dE/dt\sim1-2\times10^{44}\,\mathrm{erg}\,\mathrm{Mpc}^{-3}\,\mathrm{yr}^{-1}$ lead to optical depth for $p\gamma$ interactions as $\tau_{p\gamma} \gtrsim 0.4$, assuming that muon neutrino emission rate follows a power law $E^{-\Gamma}$, where $\Gamma\simeq2.19$ \citep{IceCube2017}. 
Using the connection between $\tau_{p\gamma}$ and $\tau_{\gamma \gamma}$ as \citep[][]{Murase2016,Halzen2019}
\begin{equation}
\tau_{\gamma \gamma} \approx \frac{\eta_{\gamma \gamma} \sigma_{\gamma \gamma}}{\eta_{p \gamma} \hat{\sigma}_{p \gamma}} \tau_{p \gamma},
\end{equation}
$\tau_{p\gamma} \gtrsim 0.4$ gives $\tau_{\gamma \gamma} \simeq \mathcal{O}(100)$, where the corresponding values of $\eta_{\gamma \gamma}$, $\eta_{p \gamma}$, $\sigma_{\gamma \gamma}$ and $\hat{\sigma}_{p \gamma}$ are given for the $\Delta$-resonance and direct pion production by e.g. \cite{Halzen2019}. High-energy pionic $\gamma$-rays are then unable to escape from an emission region that effectively generates high-energy neutrinos.

Astrophysical (median) muon neutrino energies ($E_\nu$) between $119\,\mathrm{TeV}$ and $\mathrm{4.8\,PeV}$ in the eight years of IceCube data \citep{IceCube2017} correspond to proton energies ($E_p$) from $2.4\,\mathrm{PeV}$ to $96\,\mathrm{PeV}$ (since $E_p\approx 20 E_\nu$). Assuming the meson production is dominated by the $\Delta$-resonance, these protons interact with X-ray and UV target photons \citep[see in e.g.][]{Murase2018}, and constrain the two-photon annihilation depth at $E_\gamma \sim 5$--$200\,\mathrm{GeV}$ energies. This energy range is well within the Fermi-LAT energy range of $0.1$--$300\,\mathrm{GeV}$. Considering head-on collisions, the $5$--$200\,\mathrm{GeV}$ Fermi photons produce $e^+ e^-$ pairs with $10\,\mathrm{keV}\,(\gamma/10)^2$--$3 \,\mathrm{eV}\,(\gamma/10)^2$ photons. In blazar environments the electron Lorentz factor is relativistic, rendering the energy of photons absorbing pionic $\gamma$-rays to keV, MeV or even GeV regime.

As inverse Compton scattering is Klein-Nishina suppressed at the highest energies \citep[e.g.][]{Moderski2005}, we consider that $e^+ e^-$ pairs produced in the interaction of GeV pionic $\gamma$-rays with X-ray target photons lose their energy primarily via synchrotron cooling. The characteristic frequency of synchrotron emission ($f_\nu$) by Bethe-Heitler pairs is expected in keV regime, while for photomeson production (including pairs from two-photon annihilation) $f_\nu$ is expected in the MeV regime \citep[][]{Murase2018}. Above $10^{18}$ eV the Bethe-Heitler, above $5\times10^{19}$ the photomeson pair production dominates, therefore $f_\nu$ shifts from keVs to MeVs as cosmic rays gain ultra-high energies and begin to produce neutrinos.

This scenario is consistent with multimessenger observations of the extragalactic, diffuse high-energy emission: in order to explain the diffuse high-energy neutrino flux together with the Fermi detection of the extragalactic $\gamma-$ray flux can only be done by either assuming very flat neutrino spectra or taking into account strong $\gamma$-ray absorption in the sources \citep{Ahlers2015}. 

\subsection{A case study: PKS 1502+106}

\begin{figure}
\centering
\includegraphics[scale=0.525]{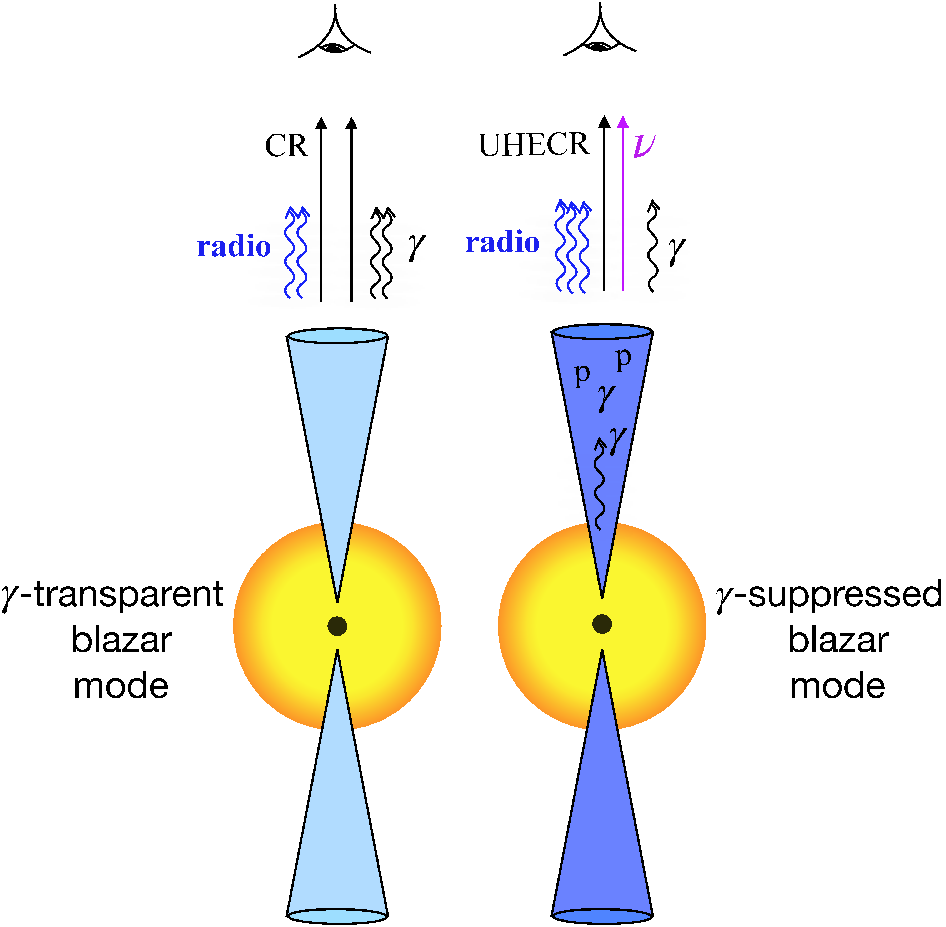}
\caption{{\bf $\gamma$-transparent and $\gamma$-suppressed mode of blazars}. In the $\gamma$-transparent blazar mode radio and $\gamma$-rays correlate with each other while cosmic-rays (CR) leave the jet. In the $\gamma$-suppressed mode temporary appearance of target photon and proton fields supply seed particles for enhanced $\gamma$-ray, CR and neutrino emission, leading to temporary suppression of the observable $\gamma$-emission due to a large $\gamma \gamma$ opacity. \label{fig:blazarsuppressed}}
\end{figure}

Observations combined with the above scenario suggest that in mode~$1$ of PKS 1502+106 $\gamma$-rays originate in a transparent source with insufficient target photon density to produce a detectable neutrino flux ($\gamma$-transparent mode). The strong correlation between the $\gamma$ and radio flux suggests the leptonic origin of $\gamma$-rays in this state. In mode~$2$ a dense photon (or even proton) target field interacting with the jet provides the conditions to produce neutrinos ($\gamma$-suppressed mode). The pionic $\gamma$-rays accompanying cosmic neutrinos lose energy in the source due to large $\tau_{\gamma \gamma}$ as shown above, and emerge at Earth below the Fermi-LAT detection threshold. We show the $\gamma$-transparent and $\gamma$-suppressed blazar modes in Fig. \ref{fig:blazarsuppressed}.

The powerful radio flare started in the beginning of 2014 reveals a very active outflow. We calculated the $\gamma$ to radio ratio in 56 time bins of the $\gamma$-suppressed mode. This ratio was the smallest in the time bin of the neutrino, yielding a p-value of $0.018$. This suggests the suppression was most effective about the IceCube neutrino detection.

A high cascading contribution from the pionic $\gamma$-rays is expected in the $\gamma$-suppressed mode, leading to a high X-ray flux. We show the Swift-XRT $0.3$--$10\,\mathrm{keV}$ light curve of PKS\,1502+106\footnote{\burl{ https://www.swift.psu.edu/monitoring/source.php?source=PKS1502+106}} in Fig. \ref{fig:lightcurves} \citep{Stroh2013}. 
We see the $0.1-300$~GeV $\gamma$-ray flux drops after a sharp peak at about MJD~57000 (Fig. \ref{fig:lightcurves}). The X-ray flux was peaking at about MJD~58100, the $\gamma$-flux being still above average, with no neutrino detected. The scenario suggests Swift observed the synchroton emission of Bethe-Heitler pairs these times. The increasing radio flux might reveal particle acceleration, protons probably co-accelerating with the electrons/positrons. IceCube detected a neutrino while the $\gamma$-flux was maximally suppressed, and the $0.3$--$10\,\mathrm{keV}$ X-ray flux was about halved. The lowered X-ray flux might be due to that at sufficiently high energies photomeson production dominates the pair production instead of the Bethe-Heitler, and a large part of the synchrotron emission from pairs appears at MeV energies, outside the $0.3$--$10\,\mathrm{keV}$ range of Swift.

This is a scenario that can work as already discussed by \citet{Halzen2020}, but to prove this in detail for PKS\,1502+106, detailed theoretical modeling of the multiwavelength data is necessary, which is beyond the scope of this paper.
\begin{figure*}
\centering
\includegraphics[scale=0.73,angle=270]{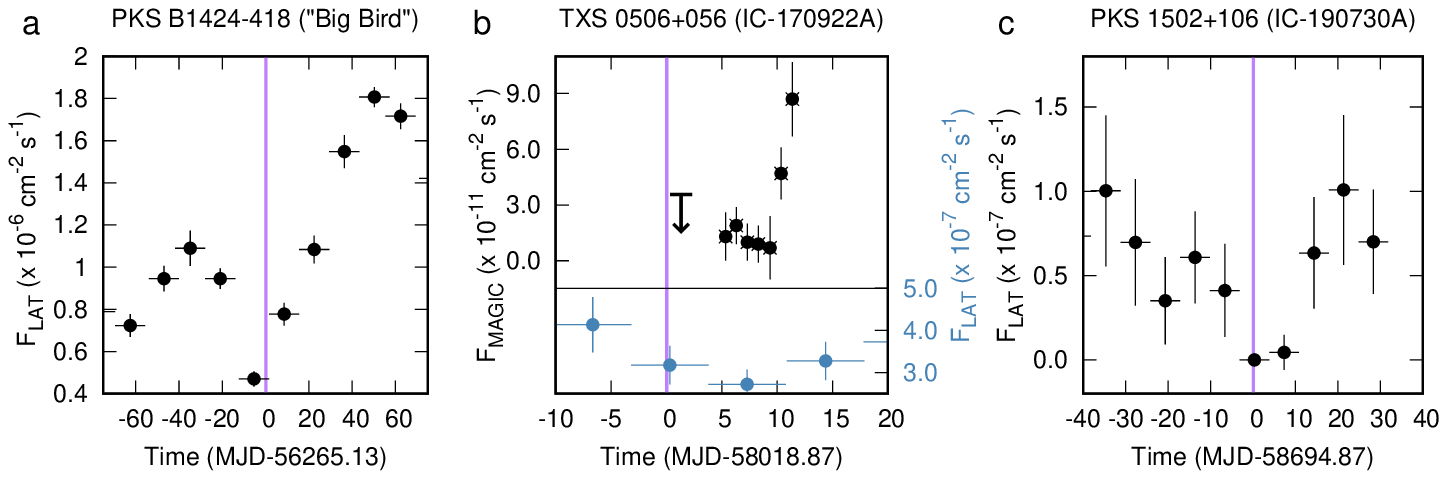}
\caption{{\bf $\gamma$-ray light curves for three blazars with coincident high-energy neutrinos.} {\bf a}: PKS\,B1424-418 as measured by Fermi-LAT (100\,MeV to 300\,GeV, $14$-days binning), plotted by black dots with error bars \citep{Kadler2016}. The vertical purple line marks the detection time of the neutrino event HESE 35. {\bf b}: The very-high energy $\gamma$-ray light curve of TXS 0506+056 as measured by MAGIC (E$>90$~GeV), plotted by black dots with error bars, and the high energy $\gamma$-ray light curve of TXS 0506+056 as measured by Fermi (E$>100$~MeV, $7$-days binning), plotted by blue dots with error bars \citep{ICTXS2018a}. The vertical purple line marks the detection time of the neutrino event IC-170922A. {\bf c}: Fermi-LAT $\gamma$-ray light curve of PKS 1502+106 (100\,MeV to 300\,GeV, $7$-days binning), plotted by black dots with error bars. The vertical purple line marks the detection time of the neutrino event IC-190730A. The arrival time of the neutrinos sets the zero value on the time axis of the plots. \label{fig:fourpiece}}
\end{figure*}

\section{$\gamma$-suppression in other blazars}

\noindent{\bf TXS\,0506+056.} IceCube recorded a $290$\,TeV muon neutrino in temporal and directional coincidence with a $\gamma$-flare of TXS\,0506+056, the correlation being statistically significant at $3\sigma$ level \citep{ICTXS2018a}. We show the blazar's $\gamma$-flux measured by Fermi-LAT and the Major Atmospheric Gamma Imaging
Cherenkov (MAGIC) Telescopes in the middle panel of Fig. \ref{fig:fourpiece}. The Fermi-LAT flux from TXS\,0506+056 drops by about 30\% from the week prior to the neutrino, and starts rising again within 2 weeks after the neutrino, in the middle of a large $\gamma$-ray flare. The MAGIC flux is undetectable at the time of the neutrino emission, but then rapidly increases by at least a factor of 10 within 2 weeks \citep{ICTXS2018a}. Interestingly, a similar rise, although on the time-scale of hours was also detected by the MASTER optical telescope in the aftermath of the neutrino detection from the blazar \citep{Lipunov2020}. We note MAGIC observed TXS 0506+056 for 2 hours on 24 September 2017 (MJD 58020) under non-optimal weather conditions \citep{ICTXS2018a}. \citet{Ansoldi2018} interpreted the $\gamma$-ray variability between MJD $58024$ and MJD $58030$ with a one-zone lepto-hadronic model, based on interactions of electrons and protons co-accelerated in a structured jet.

\noindent{\bf PKS\,B1424-418.} A temporal and directional coincidence was found between the outburst of blazar PKS\,B1424-418 and a PeV-energy neutrino event (HESE 35, dubbed as "Big Bird") \citep{Kadler2016}. In the first panel of Fig. \ref{fig:fourpiece} we show the $\gamma$-ray light curve of PKS\,B1424-418 around the time of the neutrino detection \citep{Kadler2016}. While there is a longer outburst, at the actual time of the neutrino there is a sudden suppression in the $\gamma$-flux, which then quickly rises again afterwards. Therefore, we find that the neutrino detection coincided with a local $\gamma$ minimum. \citet{Kadler2016} determined 5\% for a chance coincidence. Our scenario suggests the neutrino should come in a $\gamma$-ray dip, which significantly lowers the probability of the chance coincidence.

For comparison, we also show the $\gamma$-ray light curve of PKS 1502+106 in the third panel of Fig. \ref{fig:fourpiece}., binned in 7 days-wide bins. The temporary appearance of an X-ray/UV target photon field can temporarily suppress the $\gamma$-flux, even if the source if flaring. The time-scale of the suppression potentially depends on the characteristics of target field, which is probably more massive in case of PKS\,1502+106. To investigate origin of the target field source by source is beyond the scope of the paper.

\section{Future outlook}

We showed that the $\gamma$-ray flux of the three blazars to which a high-energy neutrino counterpart was associated, experienced a temporary $\gamma$-suppression at the time of the detected neutrino. If our scenario is correct, then it is substantially easier to discover neutrino-blazar connections due to the short allowed time coincidence between $\gamma$-suppressed periods and neutrinos. With this we suggest to revise the strategy to find the origin of the cosmic high-energy neutrinos. It is possible that previously recorded IceCube and Fermi-LAT observations are already sufficient to identify the origin of the bulk of the high-energy neutrinos detected in the Universe. 

The emerging picture from our results is that efficient neutrino production is achieved by the combination of (i) an energetic outflow from the central supermassive black hole in the blazar and (ii) the temporary increase in the target photon density. The presence of the energetic outflows are indicated by the high radio flux and the high $\gamma$-flux before and after the detected neutrinos. Therefore, $\gamma$-suppression is likely not caused by changes in the outflow from the black hole, but by the appearance of target photons at greater radii. This is a temporary condition as the $\gamma$-flux rapidly increases afterwards.

The main message of this Letter is that the temporal $\gamma$-suppression potentially resolves the apparent contradiction of the blazar models simultaneously producing a detectable neutrino flux and a $\gamma$ flare, since at the time of efficient neutrino production the observed $\gamma$-flux drops. More observations and a larger sample of sources behaving in a similar way to TXS~0506+056 and PKS~1502+106 would establish a more stable ground to this scenario, and we encourage studies in this direction.

\acknowledgments

The authors would like to thank Roger Clay and Marcos Santander for their valuable comments on the manuscript. They are grateful to the anonymous referee for the comments that improved the quality of this manuscript. E.K. was supported by the Premium Post-doctoral Research Program of the Hungarian Academy of Sciences. Support of NKFIH Grant No. 123996 in the early stage of this work is acknowledged. I.B. acknowledges the support of the National Science Foundation under grant No. \#1911796 and of the Alfred P. Sloan Foundation. JBT acknowledges support from the DFG, project TJ 62/8-1. FH's research is supported by the U.S. National Science Foundation under grants~PLR-1600823 and PHY-1913607. On behalf of Project 'fermi-agn' we thank for the usage of ELKH Cloud (formerly MTA Cloud, https://cloud.mta.hu/) that significantly helped us achieving the results published in this paper. This research has made use of data from the OVRO 40-m monitoring programme \citep{Richards2011} which is supported in part by NASA grants NNX08AW31G, NNX11A043G, and NNX14AQ89G and NSF grants AST-0808050 and AST-1109911.

\end{document}